\documentclass[11pt]{article}

\usepackage[a4paper, total={6.5in, 9in}]{geometry}
\usepackage{graphicx}
\usepackage[usenames, dvipsnames]{color}
\usepackage{shapepar}
\usepackage{picinpar}
\usepackage{amsmath}
\usepackage{amssymb}
\usepackage[square,sort,comma,numbers]{natbib}
\usepackage{lineno}

\usepackage[dvipsnames]{xcolor}

\usepackage{authblk}

\usepackage{caption}

\begin{document}

\newcommand{\SubItem}[1]{
    {\setlength\itemindent{15pt} \item[-] #1}}

\title{A sublimated water atmosphere on Ganymede detected from Hubble Space Telescope observations}

\author[$^1$]{Lorenz Roth (corresponding author, lorenzr@kth.se)}
\author[$^1$]{Nickolay Ivchenko}
\author[$^2$]{G. Randall Gladstone}
\author[$^3$]{Joachim Saur}
\author[$^4$]{Denis Grodent}
\author[$^4$]{Bertrand Bonfond}
\author[$^2$]{Philippa M. Molyneux}
\author[$^2$]{Kurt D. Retherford}

\affil[$^1$]{Space and Plasma Physics, KTH Royal Institute of Technology, Stockholm, Sweden}
\affil[$^2$]{Southwest Research Institute, San Antonio, TX 78238, USA}
\affil[$^3$]{Institut für Geophysik und Meteorologie, Universität zu Köln, Cologne, Germany}
\affil[$^4$]{Laboratoire de Physique Atmosphérique et Planétaire, STAR Institute, Université de Liège, Liège, Belgium}

\maketitle

\section*{Abstract}
Ganymede's atmosphere is produced by charged particle sputtering and sublimation of its icy surface. Previous far-ultraviolet observations of the O{\small I\,}1356-Å and O{\small I\,}1304-Å oxygen emissions were used to infer sputtered molecular oxygen (O$_2$) as an atmospheric constituent, but an expected sublimated water (H$_2$O) component remained undetected. Here we present an analysis of high-sensitivity spectra and spectral images acquired by the Hubble Space Telescope revealing H$_2$O in Ganymede's atmosphere. The relative intensity of the oxygen emissions requires contributions from dissociative excitation of water vapor, indicating that H$_2$O is more abundant than O$_2$ around the sub-solar point. Away from the sub-solar region, the emissions are consistent with a pure O$_2$ atmosphere. Eclipse observations constrain atomic oxygen to be at least two orders of magnitude less abundant than these other species. The higher H$_2$O/O$_2$ ratio above the warmer trailing hemisphere compared to the colder leading hemisphere, the spatial concentration to the sub-solar region, and the estimated abundance of $\sim$10$^{15}$ H$_2$O/cm$^{2}$ are consistent with sublimation of the icy surface as source.

\subsection*{Introduction}
%
%
Through erosion  of Ganymede's icy surface by charged particles and solar radiation, a tenuous atmosphere is formed consisting of water group molecules and atoms (H$_2$O, O$_2$, OH, H, O) \cite{johnson81,marconi07}. Molecular oxygen has long been suspected to be the most abundant constituent globally, as it does not efficiently react with the surface and is gravitationally bound. The lighter products of the ice surface erosion, H and H$_2$, escape quickly and are less abundant \cite{marconi07}. H$_2$O freezes upon contact with the ice surface, which has temperatures between $\sim$80~K and $\sim$150~K \cite{orton96,leblanc17}, also limiting the lifetime and abundance of H$_2$O in the atmosphere. Atmosphere modelling efforts suggest a dichotomy in the atmosphere between an H$_2$O-dominated atmosphere near the sub-solar point where the surface is warmest, and an O$_2$-dominated atmosphere everywhere else \cite{turc14,plainaki15,shematovich16,leblanc17}. 

The abundance of molecular oxygen and atomic hydrogen have been confirmed obervationally in several studies. Atomic hydrogen was detected through measurements of resonantly scattered solar Lyman-$\alpha$ emission in an extended corona \cite{barth97,feldman00,alday17}. First evidence for oxygen was provided by far-ultraviolet (FUV) spectra taken by the Hubble Space Telescope (HST) of oxygen emissions near 1304~Å and 1356~Å \cite{hall98}. The relative brightness of the emissions from the the spin forbidden O{\small I}($^5$S--$^3$P)~1356~Å doublet and the optically allowed O{\small I}($^3$S--$^3$P)~1304~Å triplet was used as diagnostic of the excitation process. The derived O{\small I\,}1356-Å/O{\small I\,}1304-Å oxygen emission ratio ($r_{\gamma}(\textrm{O\small{I}})$) of 1.3$\pm$0.3 was interpreted to relate to global dissociative excitation of O$_2$ by electrons. The reasoning is that electron-impact excitation processes that involve other possible species in the atmosphere such as O or H$_2$O produce substantially brighter 1304~Å emissions ($r_{\gamma}(\textrm{O\small{I}}) < 1$) \cite{doering89a,johnson03,makarov04}. In addition, resonant scattering of solar emissions by atomic oxygen contributes to the O{\small I\,}1304~Å brightness, but is absent at 1356~Å \cite{hall95}. Electron impact excitation of O$_2$, in contrast, results in a larger O{\small I\,}1356-Å/O{\small I\,}1304-Å ratio of $r_{\gamma}(\textrm{O\small{I}}) > 2$ \cite{kanik03}. The same diagnostic was used to derive O$_2$ in the atmospheres of Europa \cite{hall95,roth16-eur} and Callisto \cite{cunningham15}.

The first FUV images of Ganymede revealed that the O{\small I\,}1356~Å emissions on the orbital trailing hemisphere of Ganymede are clustered near the magnetic (and planetocentric) poles similar to an appearance of auroral bands \cite{feldman00}. Further images of the orbital leading and sub-Jovian hemispheres revealed different oxygen emission morphologies, which showed that the regions of brightest O{\small I\,}1356~Å emissions are roughly colocated with the open-closed-field-line-boundary (OCFB) of Ganymede's mini-magnetosphere \cite{mcgrath13}. On the orbital leading hemisphere which is also the plasma downstream or wake hemisphere, the band-like emissions are close to the equator, as the magnetosphere is stretched due to magnetic stresses \cite{mcgrath13,saur15}. The high-latitude emissions observed on the trailing hemisphere, which is also the plasma upstream hemisphere, are consistent with a compressed magnetosphere pushing the OCFB further to the poles \cite{musacchio17}. 

Further observations and analysis of the O{\small I\,}1356-Å/O{\small I\,}1304-Å ratio on Ganymede's trailing and leading hemispheres showed a consistently higher oxygen emission ratio ($r_{\gamma}(\textrm{O\small{I}}) \gtrsim 2$) on the leading hemisphere compared to the trailing hemisphere ($r_{\gamma}(\textrm{O\small{I}}) < 2$) \cite{feldman00,molyneux18}. The low ratio on the trailing hemisphere was interpreted to be related to a higher mixing ratio of atomic oxygen in the O$_2$ atmosphere of $\gtrsim$10\% \cite{molyneux18}, similar to interpretations of Europa's oxygen emission ratios \cite{hall95,hall98,roth16-eur}. The absolute observed O{\small I\,}1304~Å intensity together with the low oxygrn ratio requires a substantial amount of O, putting the O atmosphere in an optically thick range at O{\small I\,}1304~Å \cite{molyneux18}.

Studies of the spatial distribution were so far based almost only on the O{\small I\,}1356~Å emission, because the signal-to-noise ratio is considerably higher due to the higher emission intensity and lower background signal.  Only one recent study \cite{molyneux18} briefly discusses O{\small I\,}1304~Å images from individual exposures, mentioning large regions on the disk where $r_{\gamma}(\textrm{O\small{I}}) < 1$.

Here we present the first evidence for H$_2$O in Ganymede's atmosphere through a combined analysis of new spectra taken in 2018 by the Hubble Space Telescope's Cosmic Origins Spectrograph (HST/COS) together with archival images from HST's Space Telescope Imaging Spectrograph (STIS) from 1998 and 2010. 

First, we present two far-UV spectra taken with COS in the highest spectral resolution mode directly before and during eclipse of Ganymede by Jupiter. With this test, we show that the O atmosphere is optically thin to the 1304~Å emission, which 
sets an upper limit on the abundance of O. Analyzing the spatial distribution of the O{\small I\,}1356~Å and O{\small I\,}1304~Å emissions in STIS images, we then find that the emission ratio changes systematically with radial distance to the center of Ganymede's observed hemisphere and we show that the oxygen intensities and ratios require a substantial abundance of H$_2$O in the central sub-solar region.
\begin{table}[htb]
\caption{Parameters of the HST/COS and HST/STIS observations. CML refers to the Central Meridian (West) Longitude on Ganymede's disk, Jupiter's planetocentric longitude facing Ganymede is given by the System-III longitude.}
\label{tab:obsparam}
\resizebox*{1.\textwidth}{!}{
\begin{tabular}{lllccccccccc}
\hline
HST     & Date  & Observed  & Exp. IDs & Start	&  End  &	Total (used)& Ganymede	& Spatial   & Sub-observer	& Solar & System-III	\\
Campaign&	    & Hemisphere & / No. of   & time	&   time &    exp.time	& diameter	& resolution& CML$^*$ & phase angle	&longitude		\\ 
		&	    &           & exposures        & (UTC)	&  (UCT)    &	[s]		    & [arcsec]	& [km/pixel]       	& [$^\circ$]    & [$^\circ$]	& [$^\circ$]   \\		
\hline
\multicolumn{9}{c}{\textbf{COS observation}} \\

14634 & 2018-04-04 & sub-Jovian  & ld8k2ds1q & 14:45 & 15:16 &  1855 (0-1335) & 1.58 & n/a & 352-354 & 6.4 & 233 -- 245  \\
      &            & - eclipse - & ld8k2ds4q & 16:04 & 16:49 &  2680 (0-2160) &  & & 356-357 & 6.4 & 278 -- 298  \\
\multicolumn{9}{c}{\textbf{STIS observations}} \\

7939 & 1998-10-30 & trailing & 5 exposures & 08:21 &  13:36  & 5136 & 1.71 & 76 & 289 -- 300 & 8.6 & 226 -- 45  \\
12244 & 2010-11-19 & leading & 5 exposures & 20:11 & 03:00$^{+1}$  & 5515 & 1.64 & 79 & 98 -- 111 & 10.3 & 180 -- 43   \\
\hline
\multicolumn{9}{l}{$^*$ Central Meridian West longitude} \\
\end{tabular}
}
\end{table}

\subsection*{Results}
\begin{figure}
	\centering
		\includegraphics[width=.4\textwidth]{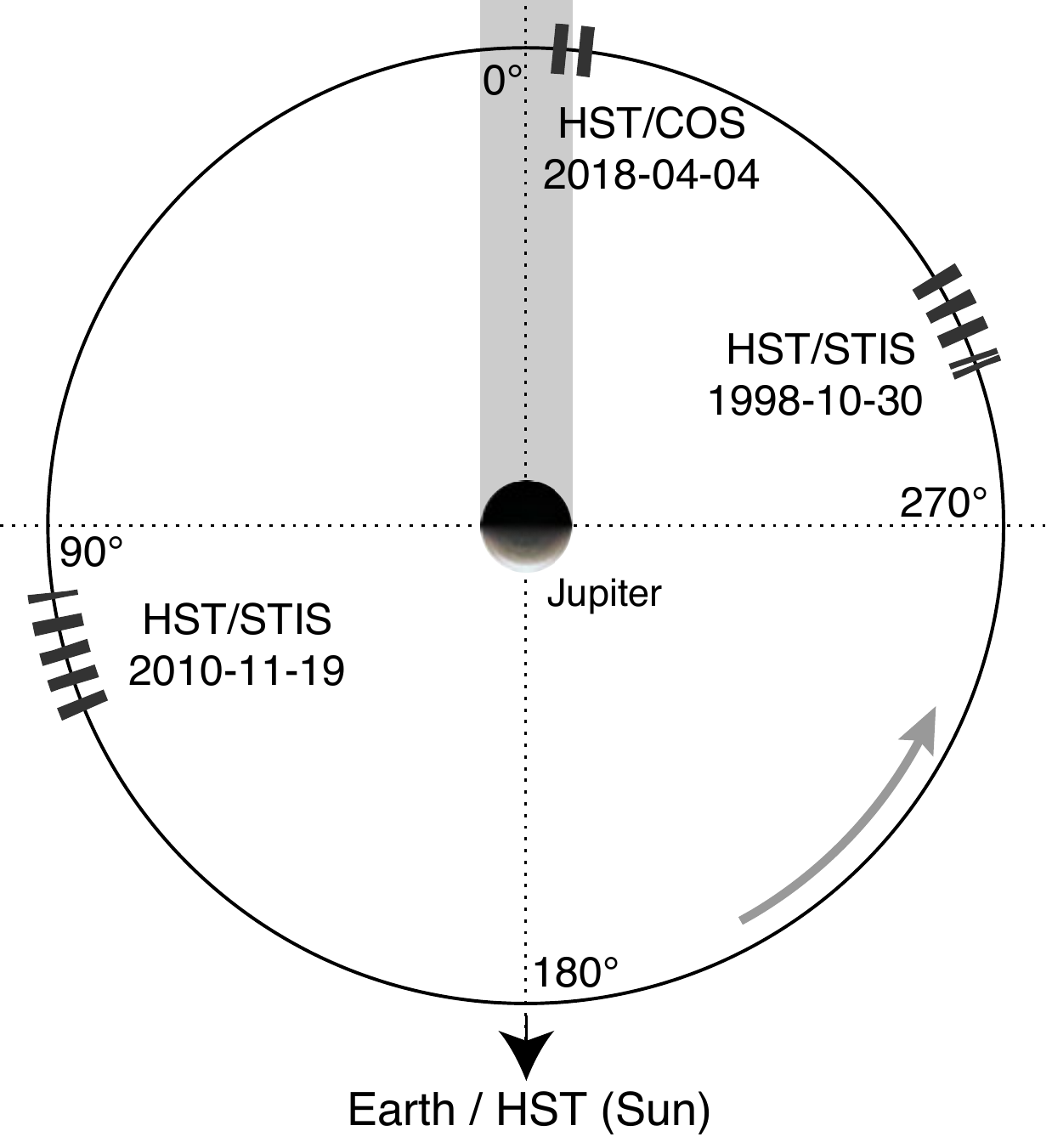}
    	\caption{Ganymede's orbital longitude (OLG) during the individual exposures of the three HST visits analyzed here. The width of each box reflects the covered orbital longitudes from start to end of an exposure. The orbit direction and also rotation of the surrounding plasma is shown by the curved arrow. (The direction to the Sun is almost identical to the Earth/HST direction, see solar phase angles in Table \ref{tab:obsparam}.)}
	\label{fig:OrbitSketch}
\end{figure}

\begin{figure}
	\centering
		\includegraphics[width=.8\textwidth]{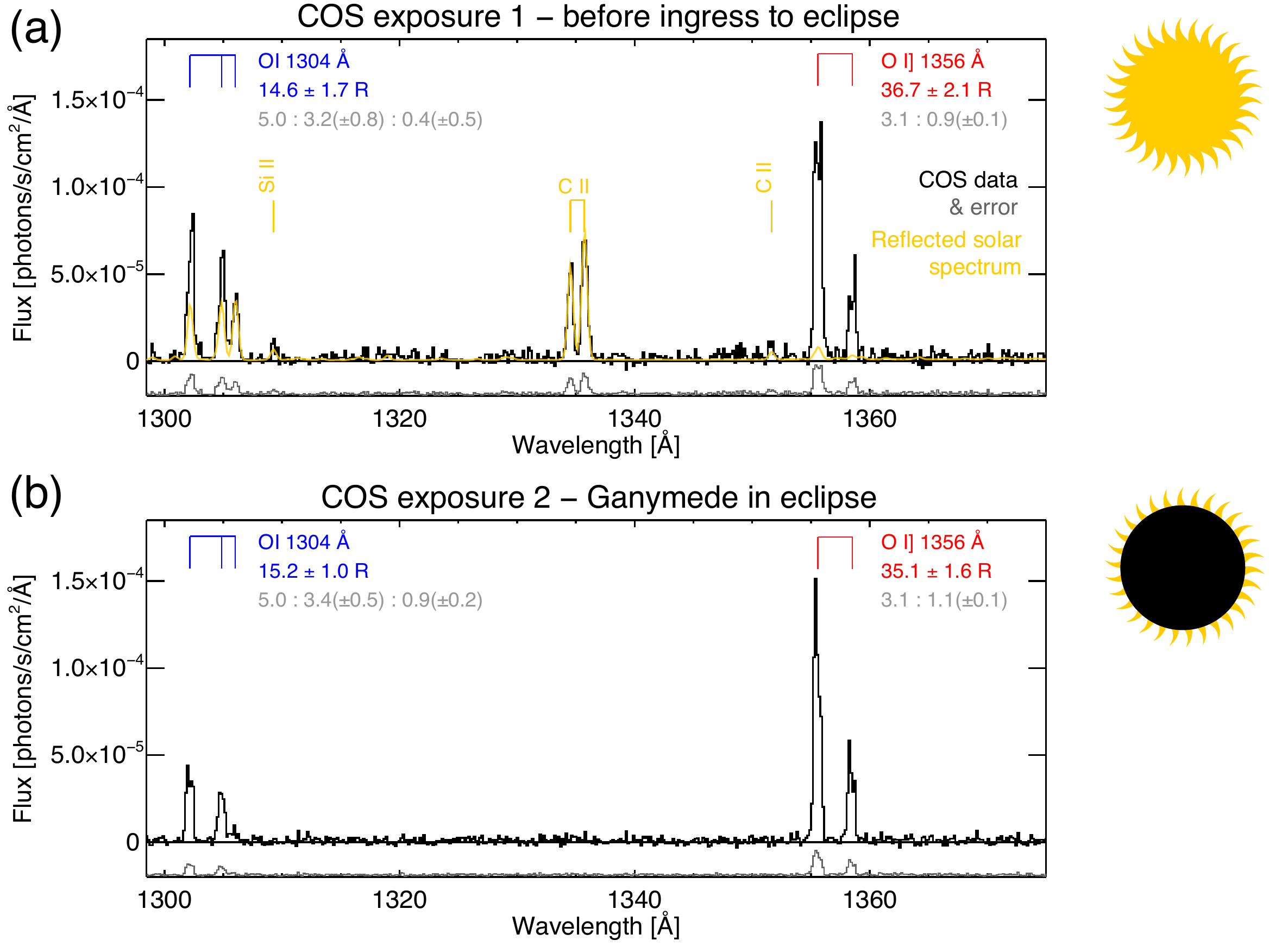}
    	\caption{COS spectra from exposure 1 (ld8k2ds1q in Table \ref{tab:obsparam}) in sunlight (a) and exposure 2 (ld8k2ds4q), when Ganymede was eclipsed by Jupiter (b). The propagated error is shown in grey, at a small negative offset on the y axis for readability. A solar spectrum (yellow) is adjusted to the sunlit exposure, with prominent solar lines indicated. The integrated intensities with propagated uncertainties are shown for the atmospheric O{\small I\,}1304~Å (blue) and O{\small I\,}1356~Å (red) emissions {(after subtraction of the solar reflection for exposure 1)}, and the relative intensities of the individual multiplet lines are given below in light grey with the brightest lines set to 5.0 and 3.1, respectively. The stability of the O{\small I} emissions in and out of eclipse rules out solar scattering as emission source and limits the O abundance.}
	\label{fig:COSspectra}
\end{figure}

COS observed Ganymede with two exposures directly before ingress into Jupiter's shadow (exposure 1) and in total umbral eclipse (exposure 2) (Figure \ref{fig:OrbitSketch}). The observational setup and data processing are explained in the Methods section. The obtained spectra are shown in Figure \ref{fig:COSspectra}.  
If solar resonant scattering by an atomic oxygen atmosphere contributes to the 1304~Å emission (in addition to electron excitation of O) as suggested in previous studies \cite{hall98,molyneux18}, the O{\small I\,}1304~Å intensity should drop from exposure 1 to exposure 2 due to the absence of scattered sunlight in the eclipse exposure. Instead of a drop, the measured O{\small I\,}1304~Å intensity hardly changed from 14.6$\pm$1.7~R (1 R or Rayleigh $= \frac{10^6}{4\pi} \frac{\textrm{photons}}{ \textrm{sr } \textrm{cm}^{2} \textrm{ s}}$) in sunlight to 15.2$\pm$1.0 R in eclipse. In the optically thin limit, the brightness of the scattered signal is the product of the O column density and photon scattering coefficient or g-factor, which is 5$\times$10$^{-7}$ photons/s for O{\small I\,}1304~Å. Setting the limit for a O{\small I\,}1304~Å decrease to $-$1 R (consistent with the +0.6 $\pm$1.0 R \textit{increase} from sunlight to eclipse within $\sim$1.5$\sigma$), we get an upper limit for the O column density of 2$\times10^{12}$~cm$^{-2}$.

The intensity of the semi-forbidden O{\small I\,}1356~Å doublet, which can only be excited by electron impact, decreases from 36.7$\pm$2.1~R (in sunlight) to 35.1$\pm$1.6~R (in eclipse). This suggests that the auroral (electron) excitation does not change or slightly decreases between the two exposures, which excludes the possibility that an increase in aurora excitation cancels out and thus masks a potential drop in resonant scattering of O{\small I\,}1304~Å. Furthermore, we have analyzed the time series of the intensities over the exposures (Extended Data Figure 1), which show that the intensities of both oxygen multiplets do not undergo systematic changes but appear to be stable throughout the exposures. This stability as well as the measured line ratios within the O{\small I\,}1304~Å triplet further support the conclusion that the resonant scattering contribution is negligible (Methods section).

The upper limit on O abundance is well below previously estimated values \cite{molyneux18} and rules out atomic oxygen as a viable emission source. We now investigate the spatial distribution of the relative O{\small I\,}1304~Å and O{\small I\,}1356~Å emissions in order to get more insights into the nature of the oxygen emission ratio and in particular the difference in the ratio between the trailing and leading hemispheres. 

\begin{figure}
	\centering
		\includegraphics[width=\textwidth]{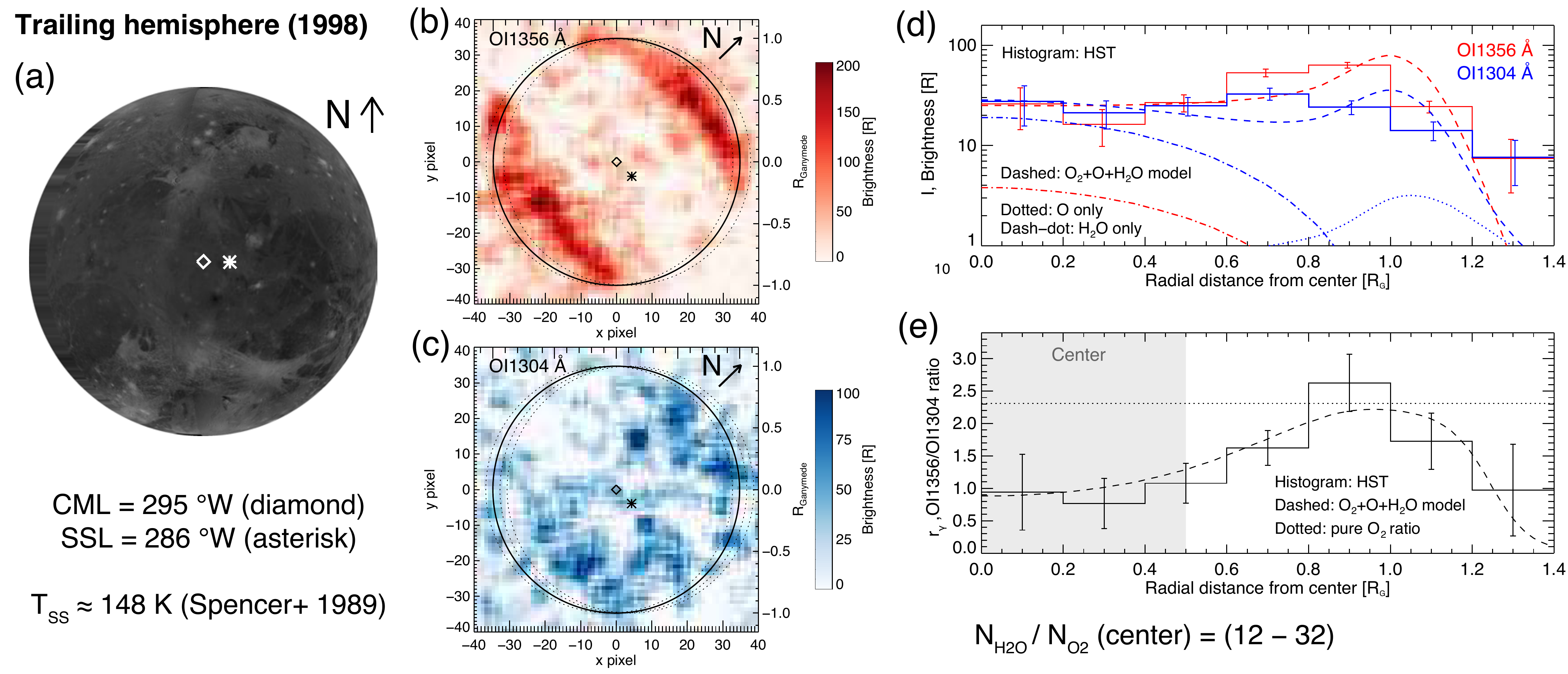}
    	\caption{Observation of Ganymede's trailing hemisphere. (a) Projection of a visible image mosaic \cite{usgs_gaymap}. (b) and (c) HST/STIS images from 1998 of the O{\small I\,}1356~Å and O{\small I\,}1304~Å emissions, respectively. The vector shows the direction to Jupiter North (N). The slightly dispersed locations of Ganymede's disk at the individual multiplet lines are shown by dotted circles. Diamonds indicate the disk center and the asterisks the sub-solar point (in a,b, and c). SSL is the observed sub-solar longitude, CML the central meridian longitude, and the given temperature at the sub-solar point of $T_{SS} = 148$~K is estimated with a thermal model \cite{spencer89}. (d) Radial brightness profiles of the average O{\small I\,}1356~Å (red) and O{\small I\,}1304~Å (blue) brightness and propagated uncertainties within concentric rings from disk center out to $\geq$1.4 R$_\mathrm{G}$. Radial emissions profiles of the simulated aurora with the assumed model atmosphere are shown for the total modelled intensity (dashed), and for the individual contributions from O (dotted), and H$_2$O (dash-dotted). The O$_2$-only model can be inferred from the differences of these models. (e) The profile of the observed oxygen emission ratio $r_{\gamma}(\textrm{O\small{I}})$ (bottom) agrees with the ratio of O$_2$ (dotted) only near the limb but is in good agreement with the O$_2$+O+H$_2$O atmosphere model (dashed) at all radial distances. The derived column density ratio $N_{\mathrm{H}_2\mathrm{O}}/N_{\mathrm{O}_2}$ in the center region (shaded grey in panel e) is shown at the bottom.}
	\label{fig:STIStrail}
\end{figure}

\begin{figure}
	\centering
		\includegraphics[width=\textwidth]{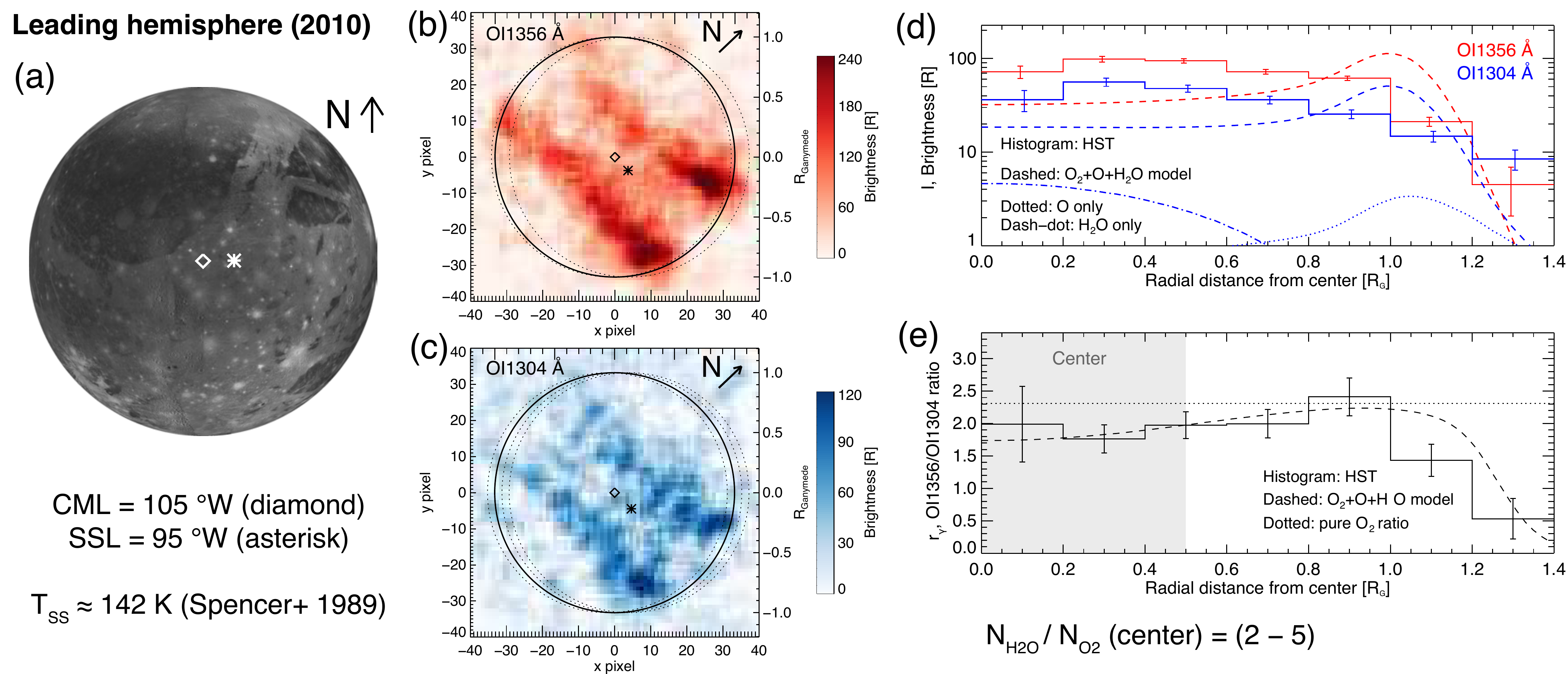}
    	\caption{Observation of Ganymede's leading hemisphere. For details on the panels see caption of Figure \ref{fig:STIStrail}. The estimated surface temperature (142~K) is lower as the albedo is higher. The measured profile of the line ratio is qualitatively similar to the trailing hemisphere but decreases only to $\sim$1.8 on the disk. The derived column density ratio $N_{\mathrm{H}_2\mathrm{O}}/N_{\mathrm{O}_2}$ (bottom) is accordingly lower as expected for the colder surface, if sublimation is the source.}
	\label{fig:STISlead}
\end{figure}

The trailing and leading hemispheres were imaged during three and four HST campaigns, respectively. The two best suitable HST visits for our analysis were selected considering two criteria: (1) the signal-to-noise ratio in the O{\small I\,}1304~Å emission, which is determined by the combined exposure time during low geocorona phase (when HST is on the Earth's night side); (2) Ganymede's angular diameter. A diameter of $\sim$1.6" is optimal for our analysis, because the 2" wide aperture slit then also captures the region above Ganymede's limb while the images provide good spatial resolution across the disk (c.f. Table \ref{tab:obsparam}). The STIS data processing and extracting of the coadded 1304~Å and 1356~Å images (Figures \ref{fig:STIStrail} and \ref{fig:STISlead}, panels b and c) are described in the Methods section. 

The O{\small I\,}1304~Å morphology generally resembles the O{\small I\,}1356~Å morphology, as found previously \cite{molyneux18}. The image-averaged ratios of $\langle r_{\gamma}(\textrm{O\small{I}})\rangle = 1.8\pm0.2$ (trailing) and $\langle r_{\gamma}(\textrm{O\small{I}})\rangle = 2.0\pm0.1$ (leading) are also roughly consistent with values previously derived from the same \cite{feldman00} or other HST data \cite{hall98,molyneux18}. 

On closer inspection, the O{\small I\,}1304~Å emissions become relatively stronger towards the disk center compared to the O{\small I\,}1356~Å emissions. This is the case on both hemispheres, but is particularly obvious on the trailing hemisphere. We therefore calculated radial intensity profiles for both oxygen emissions and their ratio, $r_{\gamma}(\textrm{O\small{I}})$, with radial bins in steps of $\Delta r$ of 0.2 R$_\mathrm{G}$ (Ganymede radius R$_\mathrm{G}$ = 2634~km) (panels d and e in Figures \ref{fig:STIStrail} and \ref{fig:STISlead})).

The \textit{emission} profiles (panels d) reflect the structure of the auroral bands: the near-equator bands on the leading hemisphere result in the highest intensities near radius $\sim$0.3 R$_\mathrm{G}$; the more polar bands on the trailing hemisphere lead to the brightest emissions closer to $\sim$1 R$_\mathrm{G}$.

The \textit{ratio} profiles (e panels), however, are similar on both hemispheres: The ratio peaks close to the limb ($\sim$1 R$_\mathrm{G}$) with a maximum value of $r_{\gamma}(\textrm{O\small{I}}) \sim 2.4$, consistent with electron-impact on a pure O$_2$ atmosphere. From these maximum values $r_{\gamma}(\textrm{O\small{I}})$ systematically decreases towards the disk center as well as towards higher distances above the limb. The mean ratios with uncertainty $\sigma$ in the disk centers ($r < 0.5$~R$_\mathrm{G}$, shaded areas in Figures \ref{fig:STIStrail}e and \ref{fig:STISlead}e) are $r_{\gamma}(\textrm{O\small{I}}) = 0.97\pm0.22$ on the trailing hemisphere and  $r_{\gamma}(\textrm{O\small{I}}) = 1.83\pm0.16$ on the leading hemisphere. These values differ from a pure O$_2$ ratio of $r_{\gamma}(\textrm{O\small{I}})$=2.3 by 6.0$\sigma$ and 2.9$\sigma$, respectively, requiring another source species. We have searched for other systematic changes in the ratio, such as between the dawn and dusk sides or polar and equatorial regions, but did not find any significant or clear trends. 

In previous studies, $r_{\gamma}(\textrm{O\small{I}})$ below 2 were interpreted as higher abundances of atomic oxygen \cite{hall98,molyneux18}. With the upper limit from the COS eclipse observation, the maximum possible abundance of O and the corresponding emission intensity is, however, insufficient to explain the observed ratios ($r_{\gamma}(\textrm{O\small{I}}) < 2$) and absolute intensities (20-80 R) near the disk center. 

\begin{figure}
	\centering
	\includegraphics[width=1.\textwidth]{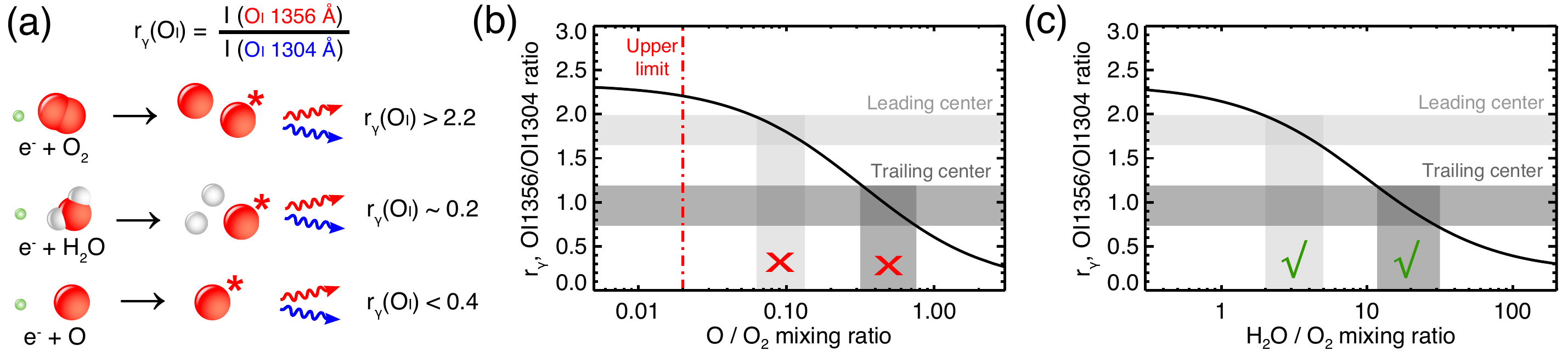}
	\caption{(a) Schematic showing how the oxygen emission ratio is diagnostic of the atmospheric composition via the relative excitation and thus emission rates \cite{doering89a,kanik01,kanik03,makarov04}. (b) and (c): O{\small I\,}1356-Å/O{\small I\,}1304-Å emission ratio as a function of mixing ratio of O and H$_2$O in an O$_2$ atmosphere for electron impact only (solid). An electron temperature of T$_e = 100$~eV is assumed but the relative exciting ratios and thus mixing ratios are similar for reasonable values between 10~eV and 200~eV. The orange and green shaded areas indicate the mean ratios $\pm$1$\sigma$ in the disk center regions ($<$0.5 R$_\mathrm{G}$) and the corresponding mixing ratio ranges. Resonant scattering of O{\small I\,}1304-Å by O is not included since its contribution was shown to be negligible in our eclipse test.}
	\label{fig:rates_ratio}
\end{figure}

The minimum O$_2$ column density in the bound atmosphere required to explain the O{\small I\,}1356~Å intensities on the disk {was shown to be} N$_{\mathrm{O}_2} = 1\times 10^{14}$~cm$^{-2}$ \cite{hall98,feldman00,molyneux18}. The upper limit on the O column density of N$_\mathrm{O} = 2\times 10^{12}$~cm$^{-2}$ derived earlier hence implies a conservative upper limit for the O/O$_2$ ratio of 0.02 in the bound atmosphere. The required O/O$_2$ mixing ratios {(see shaded areas in Figure \ref{fig:rates_ratio}b) are, however, considerably larger than 0.02 effectively ruling out atomic oxygen as the constituent that reduces $r_{\gamma}(\textrm{O\small{I}})$ in the disk centers.}


Similar to atomic oxygen, electron impact dissociation of water vapor (H$_2$O) produces considerably stronger O{\small I\,}1304~Å emissions (Figure \ref{fig:rates_ratio}a). The FUV oxygen ratio was recently used as diagnostic to distinguish between O$_2$, H$_2$O and CO$_2$ in the gas environment of comet 67P/C–G in a series of studies \cite[e.g.,][]{feldman15,feldman18} as well as to support the detection of localized H$_2$O aurora at Europa \cite{roth14-science}. Compared to O$_2$, the rate for producing the FUV oxygen emissions from H$_2$O is about an order of magnitude lower \citep{kanik03,makarov04} and the H$_2$O abundance must exceed the O$_2$ abundance to effectively reduce $r_{\gamma}(\textrm{O\small{I}})$. Indeed, H$_2$O/O$_2$ mixing ratios in the range of 12-32 (trailing hemisphere) and 2-5 (leading hemisphere) are consistent with the observed center emission ratios (see Figure \ref{fig:rates_ratio}c), implying that the atmosphere in the disk center regions is dominated by H$_2$O.

In the region above the limb ($r > 1.2$~R$_\mathrm{G}$), where $r_{\gamma}(\textrm{O\small{I}}$) also drops below 2, the absolute intensities are lower ($<$10~R) and the O$_2$ abundance is likely also lower. Here, contributions from O (and a higher O/O$_2$ mixing ratio) in the extended exosphere can possibly explain the lower oxygen emission ratio, similar to Europa \cite{roth16-eur}. Given the overall low emission intensities there, even small abundances of O near our upper limit can be sufficient to reduce $r_{\gamma}(\textrm{O\small{I}})$. The two different constituents lowering $r_{\gamma}(\textrm{O\small{I}})$ in the two different regions (H$_2$O near the center, O in the above-limb region) is also a consistent explanation for the particular radial profiles in the emission ratio, with a peak near the limb and the two separate minima.

We simulated the global electron-excited oxygen emissions for a model O$_2$-O-H$_2$O atmosphere (Methods section). The goal of the simulations is to produce emission \textit{ratio} profiles consistent with the observed profile based on simplified but reasonable neutral and electron properties.  In particular, we assumed a global O$_2$ atmosphere with a vertical column density of N$_{O_2} = 2.8\times 10^{14}$~cm$^{-2}$ everywhere and an H$_2$O atmosphere strongly concentrated around the sub-solar point.  The assumption of a global O$_2$ abundance is supported by the global presence of the 1356~Å emissions \cite{musacchio17}, which originate almost only from O$_2$ (see Extended Data Figure 2). Potential asymmetries in the global O$_2$ atmosphere, as recently proposed to be present between the dawn and dusk limb \cite{leblanc17,oza18} do not affect the oxygen emission ratio and are negligible for our analysis. The surface H$_2$O density at the sub-solar point is then adjusted to match the measured oxygen emission ratios in the disk center regions on the two hemispheres. The model line-of-sight column densities for O$_2$, O, and H$_2$O for the trailing hemisphere observation are shown in Extended Data Figure 3.  

We refrain from stating uncertainty ranges on absolute abundances assumed in the model, which would mostly reflect the uncertainty in the hardly constrained electron properties \cite{molyneux18}.  The assumed model scenario represents a reasonable O$_2$ density in Ganymede's atmosphere and a corresponding electron excitation potential consistent with the measured image-averaged oxygen intensities. Other scenarios such as a significantly higher electron excitation potential implying lower atmospheric abundances \cite{eviatar01-aur} or an overall denser atmosphere, as recently suggested \cite{carnielli20}, and lower electron excitation are also possible. Note, however, that the H$_2$O/O$_2$ ratio ranges derived above are much less sensitive to electron properties (indepedent of electron density) and thus more reliable contraints.

The simulation results show that contributions from H$_2$O and O are marginal for the total (image-averaged) O{\small I\,}1356~Å intensity, which almost entirely originates from and thus directly constrains the abundance of O$_2$ (Extended Data Figure 2). The contribution from H$_2$O and O to the O{\small I\,}1304~Å emissions averaged over the image, in contrast, are about 15\% and 10\%, affecting the oxygen emission ratio and the radial ratio profiles (Figures \ref{fig:STIStrail}e and \ref{fig:STISlead}e). 

The simulated \textit{emission} profiles (Figures \ref{fig:STIStrail}d and \ref{fig:STISlead}d) reflect mostly the limb brightening from the line-of-sight integration of the global O$_2$ atmosphere. The simulated oxygen \textit{ratio} profiles, however, rather reflect the mixing ratios and thus the H$_2$O atmosphere around the sub-solar point and the extended O exosphere at higher altitudes above the limb.   


We note that in the bins of the largest radial distance (r$>$1.2 R$_\mathrm{G}$), the observed emissions at both oxygen lines appear to be systematically higher than the vanishing emission in the simulation profiles. This difference might be related to an additional radially extended part of the exosphere \cite{eviatar01-ion,carnielli20}, which was not included here in order to keep our model simple.

The simulated O{\small I\,}1356-Å/O{\small I\,}1304-Å ratio profiles are consistent with the observed profiles on both hemispheres. The maximum is in the radial bins at or just inside 1 R$_\mathrm{G}$ in all cases, which is the region where contributions from H$_2$O and O to the O{\small I\,}1304~Å emissions (see dash-dotted and dotted lines in Figures \ref{fig:STIStrail} and \ref{fig:STISlead}) are lowest. Towards the disk center, the H$_2$O column density and hence the emission from H$_2$O increases, reducing the resulting oxygen emission ratio. At larger radial distances above the limb, the relatively higher abundance of O in the extended atmosphere leads to relatively higher O{\small I\,}1304~Å emission than O{\small I\,}1356~Å emission, and thus to the decrease of the O{\small I\,}1356-Å/O{\small I\,}1304-Å ratio.  

{Besides oxygen emission, electron impact on H$_2$O produces H~Lyman-$\alpha$ emission with an estimated intensity of about 200~R. In the avaiable data, this is, however, indistinguishable from spatially variable surface reflections at Lyman-$\alpha$ \cite{alday17}.}


\subsection*{Discussion}

The low oxygen emission ratios in the center of Ganymede's observed hemispheres are consistent with a locally H$_2$O-dominated atmosphere. With phase angles around 10$^\circ$ (Table \ref{tab:obsparam}) the disk centers are close to the sub-solar points (see diamond and asterisk in Figures \ref{fig:STIStrail} and \ref{fig:STISlead}). A viable source for H$_2$O in Ganymede's atmosphere can be sublimation in the low-latitude sub-solar regions \cite{squyres80}, where an H$_2$O-dominated atmosphere was indeed predicted by atmosphere models \cite{marconi07,plainaki15,leblanc17}. Our derived H$_2$O mixing ratios are in agreement with these predictions. While previously detected tenuous atmospheres around icy moons in the outer solar system were consistent with surface sputtering (or active outgassing) as source for the neutrals \cite{mcgrath04,teolis18,paganini19}, our analysis provides the first evidence for a sublimated atmosphere on an icy moon in the outer solar system. Water ice sublimation at heliocentric distances of $\sim$5 AU has been indirectly observed through photolysis products at comets \cite[e.g.,][]{weaver97}. {A tentative direct measurement of H$_2$O at comet C/2006 W3 at 5~AU was interpreted to origniate from sublimation from small ice-bearing grains \cite{valborro14}.}

Surface temperatures on Ganymede's dayside trailing and leading hemispheres of $\sim$148~K and $\sim$142~K are estimated by a thermal model taking into account albedo and thermal inertia of the surface \cite{spencer89}, consistent with observations \cite{orton96}. {The vapor pressure above pure ice \cite{feistel07} at these temperatures converts to H$_2$O densities of $1.7\times 10^{9}$~cm$^{-3}$ on the trailing hemisphere sub-solar point and a six times lower density of $3.1\times 10^{8}$~cm$^{-3}$ on the leading hemisphere, using the ideal gas law. This difference between trailing and leading hemisphere is remarkably similar to the difference found in our data: the emission ratios also suggest an approximately 6-fold higher H$_2$O/O$_2$ ratio above the central trailing hemisphere (Figures \ref{fig:STIStrail} and \ref{fig:STISlead}).}

{With an assumed scale height of 50~km for the modelled H$_2$O atmosphere (close to the nominal scale height at 140~K), the model surface densities are $1.2\times 10^{9}$~cm$^{-3}$ at the central trailing hemisphere and $2.0\times 10^{8}$~cm$^{-3}$ at the central leading hemisphere. These model surface densities correspond to $\sim$71\% (trailing) and $\sim$65\% (leading) of the theoretical values for pure ice stated above.} Ganymede's surface reflectance spectra were shown to be consistent with both a homogeneous surface material consisting of 90\% ice and a segregated surface, with regions of pure ice (50\% of the area) and ice-free regions covered by dark deposits \cite{spencer87_gan_cal}. Recent studies suggest that the abundance of darkening material relative to ice is higher near the equator and particularly on the trailing hemisphere \cite{ligier19,mura20}, which apparently does not or hardly impede sublimation there. The vapor density fractions of 71\% and 65\% of the values for pure ice found for the two hemispheres is suggestive of a relatively high ice fraction; the various measurements and model uncertainties (in particular the uncertainty about the electron properties and the sensitivity of the vapor pressure to small temperature changes) prevent further conclusions. Additionally, recent analysis of Ganymede's FUV reflectance suggests that the surface ice contains a small fraction ($<$1\% by volume) of UV-absorbing impurities \cite{molyneux20}, which may modify the expected sublimation rate. Sublimation and subsequent volatile transport is also likely to affect the geological landforms on Ganymede leading to, e.g., formation of specific dark features \cite{prockter98,mangold11}.

At the terminator, the temperature drops below 100~K \cite{orton96,dekleer21} and the H$_2$O vapor pressure is negligible, consistent with $r_{\gamma}(\textrm{O\small{I}})>2$ and pure O$_2$ near the limb. Thus, our results suggest that Ganymede's atmosphere posseses a pronounced day/night asymmetry. Particularly when the trailing hemisphere is illuminated there seems to be a density difference from the sub-solar point to the terminator (and night side) by more than an order of magnitude, possibly leading to atmospheric day-night winds as expected to be present at Io \cite{ingersoll85} if Ganymede's atmosphere is collisional. Such asymmetries also essentially affect the space plasma and magnetospheric environment and need to be taken into account in numerical simulations and future data analysis for a comprehensive understanding of the Ganymede system (e.g., \cite{jia09,duling14,fatemi16,carnielli20}). 

The Jupiter Icy Moons Explorer (JUICE) mission of the European Space Agency will investigate Ganymede in great detail with its eleven science instruments during several flybys in the first mission phase (expected for the period 2031-2034) and finally from orbit around the moon (from year 2034). Several science instruments are equipped to measure Ganymede's neutral gas environment and in particular the H$_2$O abundance by remote sensing of UV, optical, infrared, and sub-millimeter emissions \cite{plainaki20,wirstroem20}, as well as in-situ with the neutral particle detector. 

The JUICE mission plan is to orbit Ganymede for at least 280 days, in a variety of orbits including circular phases at $\sim$5000~km and $\sim$500~km altitude. During these periods the JUICE UV spectrograph (UVS) instrument plans to map the location, brightness, and altitude distribution of OCFB emissions of  O{\small I\,}1356~Å and O{\small I\,}1304~Å over a wide range of longitudes. For the planned surface reflectance mapping measurements at far-UV wavelengths using the 1650~Å absorption edge feature \cite{becker18,molyneux20}, it will be useful to search for latitudinal trends associated with the transport of water vapor away from the sub-solar point and to determine if the transport extends to persistently shaded regions near Ganymede's poles. 

The atmospheric and surface UVS measurements will be combined with observations of both the surface and atmosphere by other JUICE instruments, to determine the sources and sinks of Ganymede’s exosphere and ionosphere. Our results place observational constraints on the contribution of sublimation to the atmosphere, and provide the JUICE instrument teams with valuable information that may be used to refine their observation plans to optimize the use of the limited spacecraft power and telemetry resources.

\section*{Acknowledgments}
L. R. appreciates the support from the Swedish National Space Agency (SNSA) through grant 154/17 and the Swedish Research Council (VR) through grant 2017-04897. J.S. acknowledges funding from the European Research Council (ERC) under the European Union’s Horizon 2020 research and innovation programme (grant agreement No. 884711).

\section*{Author contributions}
L.~R. led the study, performed the data analysis, and wrote the manuscript. N.~I. supported all steps of the data analysis. G.~R.~G. contributed to the analysis and interpretation of the COS eclipse test. L.~R., J.~S., D.~G. and B.~B. planned and performed the 2010 and 2017 HST observations and observing strategy. All authors contributed to the interpretation of results and manuscript writing.

\section*{Competing interests}
The authors declare that they have no competing financial interests.

\section*{Methods}

\subsection*{Data and Processing}

\subsubsection*{HST/COS spectra}

The two analyzed HST/COS exposures were obtained in April 2018 (Table \ref{tab:obsparam}) during two consecutive HST orbits with the G130M grating and a central wavelength of 1291~Å. The wavelength range of 1290–1430~Å on the COS detector segment A includes the emissions from the oxygen triplet around 1304~Å and the doublet at 1356~Å. The individual multiplet lines (1302.7, 1304.9, 1306.0~Å / 1355.6, 1358.8~Å) are spectrally resolved at a nominal resolution of 0.17~Å per resolution element. We discuss the measured multiplet line ratios briefly at the end of this section. Due to acquisition at the start of the first HST orbit, exposure 1 (ld8k2ds1q) is shorter, while the entire second HST orbit was used for science exposure 2 (ld8k2ds4q) (Table \ref{tab:obsparam}). Between the first and second exposure, Ganymede entered the umbral shadow of Jupiter (at 15:27 UTC on 2018-Apr-04). 

Ganymede appears to be well centered in the 2.5"-wide aperture as revealed by a Gaussian fit to the integrated signal along the spatial (y-) detector axis in the first exposure. The fit to the signal, which is dominated by the reflected sunlight, has the peak near the nominal center and a full width of 1.7", close to Ganymede's angular diameter (Table \ref{tab:obsparam}). As the pointing of HST is stable, Ganymede should be similarly centered in the aperture in the second exposure. 


Extended Data Figure 1 shows the temporal evolution of the detector count rates around the two oxygen emissions at 1304~Å and 1356~Å. Towards the end of both exposures, the O{\small I\,}1304~Å signal increases sharply due to scattered light from the Earth's oxygen geocorona when HST approaches the terminator region. We hence reduced the exposure times by 520~s each (Table \ref{tab:obsparam}). Apart from the O{\small I\,}1304~Å increase due to the geocoronal signal, the count rates seem to be stable for both lines and in both exposures. The systematically higher O{\small I\,}1304~Å count rate in exposure 1 when compared to exposure 2 originates from the surface reflection in sunlight, which are removed in a subsequent processing step. 

The detected counts are then integrated along the spatial axis and over the reduced exposure time to obtain a spectrum for each exposure. Using the tabulated detector sensitivity the count rate (counts / exposure time) is converted to photon flux, shown in Figure \ref{fig:COSspectra}. The propagated uncertainties that include the Poisson noise and uncertainties in the correction of the background signal and  (in the case of exposure 1) subtracted solar flux are shown per spectral bin in grey. The complete spectrum of the first exposure (top panel) contains several solar lines, with the C{\small {II}} doublet near 1335~Å being most prominent. For the removal of the reflected sunlight, we apply the same method and data as used and discussed in previous work \cite{cunningham15,roth16-ceres}. The adjusted and fitted spectrum from the Solar Radiation and Climate Experiment/Solar Stellar Irradiance Comparison Experiment (SORCE/SOLSTICE) is shown in orange and matches well both the C{\small{II}} and Si{\small{IV}} lines. As Ganymede is in total umbral eclipse throughout the second exposure, surface reflections are not present. 

We calculate total intensities measured at O{\small I\,}1304~Å and O{\small I\,}1356~Å lines and normalize them to the area of Ganymede's disk (in analogy to \cite{hall98}). 
The resulting disk-averaged intensities are shown in Figure \ref{fig:COSspectra} along with the relative intensities of the three (O{\small I\,}1304~Å) and two (O{\small I\,}1356~Å) multiplet lines. The obtained intensities are identical in the two exposures for both oxygen emissions within their respective propagated uncertainties. The resulting O{\small I\,}1356/O{\small I\,}1304 ratio is slightly higher in the first ($r_{\gamma}(\textrm{O\small{I}})=2.5\pm0.3$) than in the second exposure ($r_{\gamma}(\textrm{O\small{I}}) = 2.3\pm0.2$), although this difference is not significant either.

\subsubsection*{Oxygen multiplet line ratios in COS data}

The COS data also provided the first resolved measurement of the multiplet lines of both oxygen emissions. The multiplet ratios are diagnostic for the source processes and optical thickness effects and we therefore discuss the results here briefly.

The ratio of the three multiplet lines of the 1304~Å emissions is in both exposures (Figure \ref{fig:COSspectra}, light grey) consistent with the theoretical relative line strength for a collisionless and optically thin atmosphere. The $^3$S--$^3$P transitions have different ground states but originate from the same excited upper state (3S, J=1), and the ratios in the optically thin case relates to the Einstein A coefficients, i.e., 1302.2~Å:1304.9~A:1306.0~Å = 5:3:1 \cite{morton03}. If there is a source of emission at high optical depth in a very optically thick medium (as for the chromosphere and transition region of the Sun), then the multiple photon scattering cancels out the effect leading to similar or even slightly reversed relative intensities for the three lines \cite{gladstone92}.

In a marginally optically thick medium, there might be small deviations from the 5:3:1 ratio. Since resonance scattering involves partial frequency redistribution, there is a different scattering phase function for each line in the triplet, depending on the relative weight of Rayleigh phase function. This can lead to a change in the line ratio for a directional source flux as for the solar flux \cite{chandrasekhar60}. For the small solar phase angle during the COS observations of 7$^\circ$, a slightly different multiplet ratio of 1302.2~Å:1304.9~A:1306.0~Å = 5:3.4:1.5 is theoretically expected \citep[chapter 19 of][]{chandrasekhar60} for only solar resonant scattering. Hence, the very faint line at 1306.0~Å in the sunlit exposure (see multiplet ratio in Figure \ref{fig:COSspectra}b) would not be consistent with a dominating resonant scattering signal. However, this is only a weak diagnostic, as the differences in the multiplet ratio for resonant scattering and (isotropic) electron excitation are small.  

The obtained ratios for the O{\small I\,}1356~Å doublet (Figure \ref{fig:COSspectra}, light grey) are similarly in agreement for both exposures with the theoretical ratio of 1355.6~Å:1358.0~Å = 3.1:1.0 \citep{morton03}.

\subsubsection*{HST/STIS spectral images}

Far-ultraviolet spectral images with the G140L grating and the 52"x2" slit that include maps of Ganymede's oxygen emissions were taken within six HST campaigns (IDs 7939, 8224, 9296, 12244, 13328, 14634). We analyze the spatial distribution of the O{\small I\,}1304~Å and O{\small I\,}1356~Å emissions with image pairs from two representative HST visits for the trailing and leading hemispheres, selected after the criteria as described in the main text.

We combined all low-geocorona exposures - five from the 1998 trailing hemisphere visit (ID's o53k01010, o53k01020, o53k01040, o53k01060, o53k01080) and five from the 2010 leading hemisphere visit (IDs objy03020, objy03040, objy03060, objy03080, objy030a0) - for a superposition image with improved signal-to-noise ratio. We then follow a standard processing pipeline for correcting the detector images for background and surface reflection signals, see \cite{roth14-io} or \cite{roth16-eur} for details. We have updated the method for subtracting solar surface reflections in two aspects: because lower resolution spectra slightly smear the spectral trace, we use the high-resolution spectra from the SORCE/SOLSTICE instrument \cite{mcclintock05} and adjusted them to the STIS G140L resolution (for the 1998 visit only a UARS/SOLSTICE spectrum with lower resolution is available). Furthermore, we took into account the longitudinal variation of the solar flux for the selection of the UARS/SOLSTICE data. We used the SOLSTICE spectrum from the day closest to the HST observation day but when the solar longitude facing Earth matched the Jupiter-facing longitude from the day of the observation.

After correction for background and solar reflection, {80$\times$80} pixel images (over the full slit width) containing the oxygen emission images centered on the spectral axis at 1303.5~Å and 1356.3~Å are extracted from the spectral detector images and converted to units Rayleigh (R). The analysis is carried out in the native detector frame and original pixel resolution without smoothing. For the displays in Figures \ref{fig:STIStrail} and \ref{fig:STISlead}, the images were smoothed with a 5x5 pixel boxcar function but not rotated to a common frame. All analysis was carried out with the original data (no binning or smoothing applied). The propagated uncertainties that include the Poisson noise and uncertainties in the correction of the background signal and subtracted solar flux (in the case of exposure 1) are shown per spectral bin in grey. Statistical uncertainties based on Poisson statistics in each image pixel are propagated through this processing (i.e., the uncertainties from the correction of the background signal and subtracted solar flux added) and converted to unit Rayleigh.

We calculated image-average emission intensities taking into account all pixels within 1.25~R$_\mathrm{G}$ around the disk center and normalizing it to the area of these pixels (i.e., not to the area of Ganymede's disk as done for spectral observations without spatial information). These image-averaged intensities are given in Extended Data Figure 2 along with the modelled intensities from each atmospheric constituent and the total atmosphere.   

\subsection*{Modelling}

The goal is to reproduce the emission intensities as well as the radial line ratio profiles with a simple model with as few assumptions and parameters as possible for both the atmospheric distribution and the electron properties.

For the neutral gas distributions, we assume a global, exponentially decreasing atmospheric O$_2$ density with a fixed scale height of 100 km and a surface density of n$_{\mathrm{O}_2,0}$. This scale height is on the order of the spatial resolution of the STIS images (Figure \ref{tab:obsparam}), meaning that the density above the limb as seen by STIS decreases to 1/e of the maximum over 1-2 pixels. As discussed above, the global O$_2$ abundance is supported by the fact that OI1356~Å emissions are observed across all longitudes. The assumption is also justified by the long life time or residence time of O$_2$ in the atmosphere, which is on the order of Ganymede's orbital period \cite{marconi07,leblanc17}.

Atomic oxygen O is likely produced primarily through dissociation of the molecular atmosphere (with excess energy), suggesting a higher temperature and a more gradual radial decrease than for mainly surface-derived species. We thus assume a larger scale height of 400 km. With the derived column density upper limit, we get a fixed surface density for O of $n_{\mathrm{O},0} = 5\times 10^{4}$~cm$^{-3}$. 

As a third constituent, we assume an H$_2$O abundance from sublimation concentrated around the disk center (sub-solar point) given by
\begin{equation}
    n_{\mathrm H_2\mathrm O}(h,\alpha) = n_{\mathrm H_2\mathrm O,0} \, \cos^6(\alpha) \, \exp\left( -\frac{h}{H_{\mathrm H_2\mathrm O}} \right) \; ,
\end{equation}
where $n_{\mathrm H_2\mathrm O,O}$ is the density at the surface, $h$ the altitude above the surface, $\alpha$ the angle to the disk center, and $H_{\mathrm H_2\mathrm O}$ the scale height. The nominal scale height for H$_2$O at Ganymede's surface at T = 140K is 48 km and we hence assume a scale height of 50~km. (Due to the concentration of H$_2$O near the disk center and the absence of H$_2$O near the observed limb, the H$_2$O column density derived from the emission intensity is essentially independent of the scale height.) The approximated cosine to the sixth dependence on the angle to the sub-solar point ($\alpha$) roughly reproduces the steep gradient near ($\alpha =45^\circ$) found in models (see, e.g., figure 3 of \cite{marconi07}). 
We then produce column density maps for each of the three species by integrating along the line-of-sight with the sub-solar point on the disk center, see Exteded Data Figure 3.

To compute the emission intensities, we then need to make assumptions on the electron population that excites the emissions. Ganymede's aurora is suggested to be excited by an accelerated population of electrons with $T_e$ between 75~eV and 300~eV \cite{eviatar01-aur}. There are no measurements or independent constraints available on $T_e$ in Ganymede's environment. For our model estimations, we assume one Maxwellian electron population at 100~eV to excite the oxygen emissions (which is also the temperature where the 1356~Å yield from H$_2$O was measured \cite{makarov04}).

For the density of the electrons, we consider the extent of the auroral band emissions seen on the two hemispheres and the related total intensities. For the trailing hemisphere observations, we assume an electron density of 20~cm$^{-3}$ to roughly match the observed image-averaged OI1356~Å intensity at the assumed O$_2$ abundance. On the leading hemisphere, the observed auroral intensities are higher due to the long equatorial aurora bands, and we set the electron density to 30~cm$^{-3}$, again matching the observed image-averaged OI1356~Å intensity. These densities are slightly higher than the electron density in Jupiter's magnetosphere near Ganymede \cite{kivelson04} but lower than the peak densities measured by Galileo near closest approach to Ganymede \cite{eviatar01-ion}.

The local volumetric emission from electron (dissociative) excitation is calculated by multiplying the local neutral density with the constant electron density and emission rates. The emission rates are derived as an integral over the Maxwell–Boltzmann distribution, the electron velocity and the energy-dependent cross sections for the collision of the exciting electrons with the neutral species. Cross sections are taken from laboratory measurements of the considered species \citep{doering89a,kanik03,johnson03,makarov04}. For atomic oxygen, contributions from cascades from higher states to the upper levels are approximated. We note that electron impact on H$_2$O can produce excited neutral oxygen atoms via several dissociation channels \citep{makarov04}. Finally, the 2D emission pattern is given by the line-of-sight integral over the local intensities.

Given that acceleration processes are likely required to produce the observed emissions \cite{eviatar01-aur}, the electron distribution can be highly inhomogeneous with regions of hot (accelerated) electrons and regions of electrons with lower temperature (but possibly higher density). Our assumption of a homogeneous plasma follows previous approaches \cite{hall98,feldman00} and the results for the relative H$_2$O abundance should be relatively insensitive to the exact electron properties. Dissociative excitation of O$_2$ consistently produces an emission ratio of $r_{\gamma}(\textrm{O\small{I}}) > 2.2$ in the range of possible electron temperatures, and the emission ratios for excitation of O and H$_2$O are similarly consistently $r_{\gamma}(\textrm{O\small{I}}) << 1$. Hence, while the absolute neutral abundances depend on the loosely constrained electron properties, the required relative abundances and thus mixing ratios of O and H$_2$O do not strongly depend on the assumed electron temperature or density.  

Resonant scattering by O is considered using the estimations from \cite{cunningham15} (see their figure 9). The line-of-sight O column density near the limb locally exceeds the nominal derived upper limit, but the average column density and thus scattering contribution is $\sim$1 R in both cases thus in agreement with the COS results.

The simulated aurora images are degraded to the STIS spatial resolution and smoothed to account for the STIS point-spread-function and the offset of the individual multiplet lines from the line center (for details on the production of synthetic STIS images see our earlier work \cite{roth17-io}). We then adjust the surface densities for H$_2$O to match the OI1304~Å intensities and resulting line ratios in the observations. 

From the simulated and degraded images, we then produce radial profiles for the oxygen emissions and for the resulting oxygen emission ratio, in analogy to the derivation of the observed profiles but with smaller radial bins.

\subsection*{Data availability}
All used Hubble Space Telescope data are publicly available at the Mikulski Archive for Space Telescopes (http://archive.stsci.edu/hst/). Source data are provided with this paper.

\subsection*{Code availability}
Advanced numerical code is not used in this observational study. Information on data processing and analysis implementation is available from the corresponding author on reasonable request.

\newpage




\setcounter{figure}{0}
\renewcommand{\figurename}{Extended Data Figure}

\begin{figure}
	\centering
		\includegraphics[scale=.45]{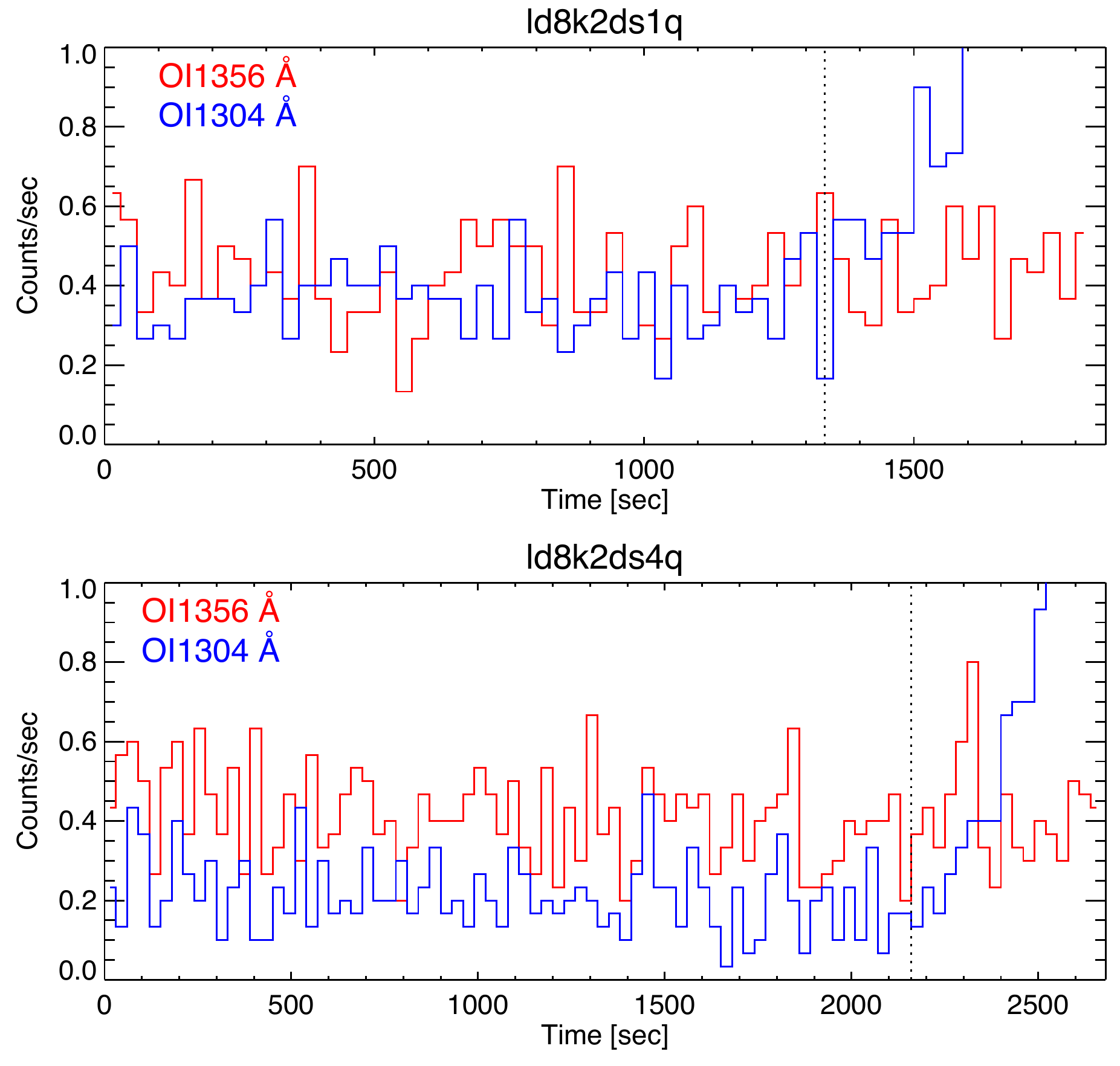}
	\caption{Count rates at the two oxygen multiplets as function of time in the two COS exposures. A sharp increase of the O{\small I\,}1304~Å emission (blue) from light scattered in the geocorona can be seen towards the end. No increase due to geocoronal scattered light is present at 1356 Å (red). For the analysis the last 520 s in each exposure are removed in the processing and only the counts left of the vertical dotted lines are used.}
	\label{fig:COStimetag}
\end{figure}

\setcounter{table}{0}
\renewcommand{\tablename}{Extended Data Table}

\begin{figure}
\resizebox*{1.\textwidth}{!}{
\begin{tabular}{lcccccccccccccccc}
\hline
Species   & Electrons & \multicolumn{3}{c}{O$_2$}    & \multicolumn{3}{c}{O}    &  \multicolumn{3}{c}{H$_2$O} & & \multicolumn{2}{c}{Total} & \multicolumn{2}{c}{Image-averaged} & Image-av.  \\		
Scale height  &  --    & \multicolumn{3}{c}{H$_{\mathrm{O}_2}=100$~km}  & \multicolumn{3}{c}{H$_\mathrm{O}=400$ km} &  \multicolumn{3}{c}{H$_{\mathrm{H}_2\mathrm{O}}=50$~km} & & \multicolumn{2}{c}{model} & \multicolumn{2}{c}{STIS intensity} & ratio \\
            & n$_e$  & n$_0$ & I$_{1356}$ & I$_{1304}$      &  n$_0$ & I$_{1356}$ & I$_{1304}$  & n$_0$ & I$_{1356}$ & I$_{1304}$ & &  I$_{1356}$ & I$_{1304}$ &  I$_{1356}$ & I$_{1304}$ & r$_{\gamma}(OI)$	\\		 
            & [cm$^{-3}$]  & [cm$^{-3}$] & [R] & [R]      &  [cm$^{-3}$] &[R] & [R]  & [cm$^{-3}$]  & [R] & [R] & & [R] & [R] & [R] & [R] & 	\\	
\hline
Trailing    & 20    & 2.8e7  & 41.3 & 17.1  & 5e4 & $<$0.1 & 2.4 & 1.2e9$^*$  & 1.0 & 5.0 & & 42.3  & 24.5 & 42.8$\pm$2.0   & 23.5$\pm$1.9 & 1.8$\pm$0.2 \\	
Leading     & 30    & 2.8e7  & 58.1 &  24.1 & 5e4 & $<$0.1 &  2.6 &  2.0e8$^*$   &  0.4 &   1.8 & & 58.5 &  28.5 & 59.0$\pm$1.7 &  30.2$\pm$1.4 & 2.0$\pm$0.1 \\	
\hline
\multicolumn{10}{l}{$^*$ maximum at the disk center / at sub-solar point.} 
\end{tabular}
}
\caption{Model atmosphere parameters and results for the corresponding oxygen intensities. The temperature of the electrons is assumed to be T$_e = 100$~eV. Maps of the model atmospheres are shown in Extended Data Figure 3. Note that the O$_2$ atmosphere produces the vast majority of the OI1356 Å emissions. For the OI1304 Å emissions, in contrast, O$_2$, O and H$_2$O all have revelant contributions to the signal. }
\label{tab:atmo_res}
\end{figure}

\clearpage

\begin{figure}
	\centering
		\includegraphics[width=\textwidth]{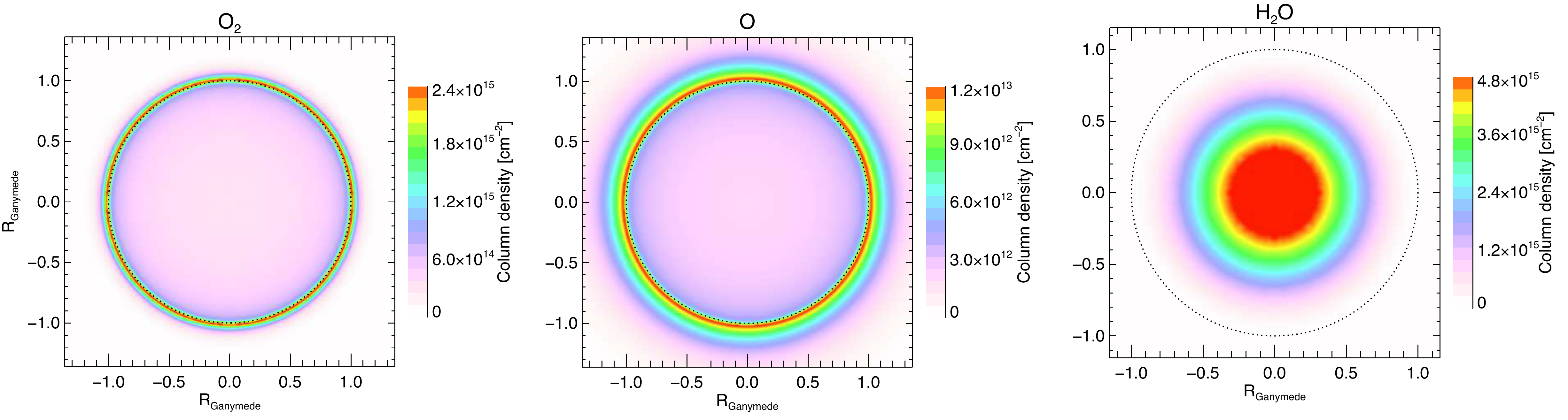}
	\caption{Column density maps of the model O$_2$, O and H$_2$O atmospheres. The H$_2$O atmosphere is scaled for the best-fit on the trailing hemisphere. The O$_2$ and O atmosphere are assumed to be identical on the trailing and leading hemispheres. The found H$_2$O density for the leading hemisphere is lower by a factor of $\sim$6.}
	\label{fig:model_atmos}
\end{figure}

\end{document}